\documentclass[apl,twocolumn,showpacs,preprintnumbers,amsmath,amssymb]{revtex4}

\usepackage{graphicx}

\begin{document}

\title{Semiconducting chains of gold and silver}
\author{Frederico Fioravante}
\affiliation{Departamento de F\'{\i}sica,  Universidade Federal de Minas
Gerais, CP 702, 30123-970, Belo Horizonte, MG, Brazil}

\author{R. W. Nunes}
\email{rwnunes@fisica.ufmg.br}
\affiliation{Departamento de F\'{\i}sica,  Universidade Federal de Minas
Gerais, CP 702, 30123-970, Belo Horizonte, MG, Brazil}

\date{\today}


\begin{abstract}
The authors introduce a geometry for ultrathin Au and Ag wires
that {\it ab initio} calculations indicate to be more stable than
previously considered planar geometries for these systems, by about
0.1 eV per atom. This structure is insulating for both metals and
for related Ag$_{0.5}$Au$_{0.5}$ alloys, with gaps of 1.3 eV for Au,
0.8 eV for Ag, and varying between 0.1 eV and 1.9 eV for the
alloys. The insulating nature of the geometry is not a result of
Peierls instabilities, and is analyzed in terms of an interplay
between geometric and electronic structure effects.
\end{abstract}

\maketitle

Nanowires (NWs) based on $4d$ and $5d$ metals are a topic of intense
current interest in the physics of nanomaterials. Understanding the
connection between atomic structure and transport, in the limit of the
ultrathin monoatomic wires that have been produced
experimentally,~\cite{ohnishi,yanson} is crucial to the future
manipulation of metallic wires and electric contacts in nanoscale
electronic devices. Previous experimental and theoretical works on
pure and alloy NWs have established that structure and transport
are dynamically and strongly linked in the nanoscale
limit.~\cite{ugarte1,ugarte2,ugarte3,liqin,enomoto} Detailed
investigation of the structural landscape of ultrathin pure and alloy
NWs is thus of paramount importance.

Previous theoretical studies
~\cite{liqin,springborg1,springborg2,ribeiro,bahn,portal} of infinite
monoatomic Au and Ag NWs have considered two planar zigzag
configurations, with angles between two noncolinear bonds along the
chain of $\sim$$60^\circ$[see Fig. 1(b)] and $\sim$$130^\circ$,
respectively. Both zigzag geometries are stable for Au, while for Ag
only the $60^\circ$ geometry is a minimum of the $E\times\ell$
surface.~\cite{ribeiro} This difference between Au and Ag has been
tied to a stronger relativistic effect in Au, that leads to enhanced
$sd$ hybridization and stabilization of the low-coordinated structure,
an effect that is also observed in the surface reconstruction of $4d$
and $5d$ metals.~\cite{smit} For both metals, the zigzag structures
remain metallic, and their stability has been connected with a gapless
Peierls transition (GPT), where transversal distortions lower the
energy, but a gap does not open at the Fermi level.~\cite{gpt} In the
recent work of Cheng and collaborators, the pathway of the thinning
process of Ag NWs, in mechanical break junction experiments
(MBJE),~\cite{yanson,ugarte1,ugarte2,ugarte3} is connected to the
successive stable minima of the $E\times\ell$ curve, as the NW radius
decreases.~\cite{cheng}
\begin{figure}
%\begin{figure}[htbp]
\includegraphics[width=8.5cm]{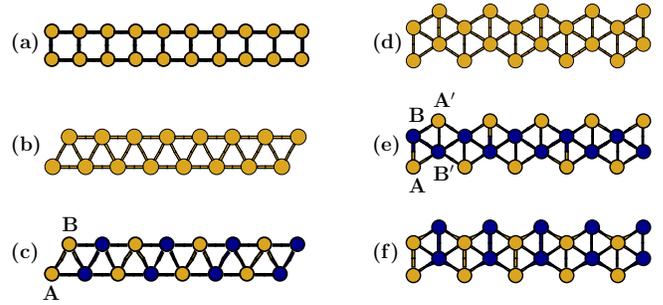}\\
\caption{(Color online) Ag, Au, and
Ag$_{0.5}$Au$_{0.5}$ nanowire structures: (a)
SQ3; (b) ZZ4; (c) A$_1$ZZ4; (d) ZZ3$+$5; (e) A$_1$ZZ3$+$5; and (f)
A$_2$ZZ3$+$5. For the alloys, light (yellow online) and dark (blue
online) circles represent Au and Ag atoms, respectively.}
\label{fig1}
\end{figure}

In the present work we examine the structure of ultrathin noble-metal
NWs. Our calculations indicate the existence of a planar
structure for both Ag and Au, and also for Ag$_{0.5}$Au$_{0.5}$
alloys, that has not been addressed in previous studies for these
systems.  In this geometry, the average coordination remains fourfold,
like in the $60^\circ$ zigzag one, but the four atoms in the unit cell
alternate between threefold- and fivefold-coordinated sites, in a
double zigzag structure [shown in Figs.~\ref{fig1}(d)-(f)] that is
lower in energy than the $60^\circ$ zigzag. Moreover, the structure is
insulating, with energy gaps of 1.3~eV and 0.8~eV for Au and Ag,
respectively, and does not result from either a Peierls or a GPT
instability. Gaps for four possible related alloy geometries vary
between 0.1 eV and 1.9 eV. No Peierls instabilities are expected for
this structure, given its insulating nature. This is a striking
result: the structure, besides being lower in energy than the planar
ones considered in previous works, is an equilibrium (unstrained)
insulating chain for these atoms, not related to Peierls
instabilities, with sizeable energy gaps.

We employ an {\it ab initio} methodology implemented in the SIESTA
package,~\cite{siesta} based on the Kohn-Sham formulation of
density functional theory (DFT).~\cite{ks} The generalized gradient
approximation (GGA)~\cite{pbe} is used for the exchange-correlation
energy, and norm-conserved Troullier-Martins
pseudopotentials~\cite{tm} represent the ionic cores. Cutoff
constrained pseudo-atomic local orbitals form the basis of
representation of the electronic wave functions, with energy shifts of
$\sim$7~mRy. A double-zeta expansion is used for each radial basis
function, including polarization orbitals for the $(l=0)~s$
functions. In all our calculations, convergence of total energy
differences and atomic forces to within 1~meV/atom and
0.003~eV/\AA, respectively, was enforced.
\begin{figure}
%\begin{figure}[htbp]
\includegraphics[width=8.5cm]{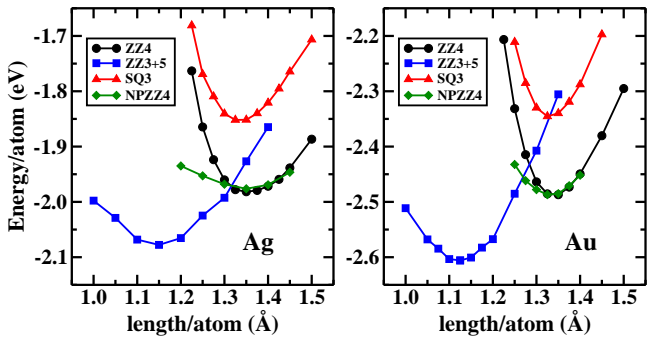}\\
\caption{(Color online) Energy (in eV/atom) $\times$ strain (in
\AA/atom) curves for Au and Ag nanowires in ZZ3$+$5, ZZ4, NPZZ4, and
SQ3 geometries.}
\label{fig2}
\end{figure}

The structures of the infinite chains we consider are shown in
Fig.~\ref{fig1}. (a) shows a planar threefold-coordinated square chain
(SQ3). (b) shows the previously-considered $60^\circ$ planar zigzag
fourfold-coordinated chain (ZZ4). (d) shows the planar double-zigzag
chain we introduce in this study, with threefold- and
fivefold-coordinated atoms (ZZ3$+$5). The latter is composed of two
intercalated sub-units, as indicated in Fig.~\ref{fig1}(e) by primed
and unprimed letters. Sites A and A$^\prime$ (B and B$^\prime$) are
both threefold (fivefold) coordinated, and are related by an inversion
symmetry in the ideal structure. We considered four alloy structures
derived from ZZ3$+$5, with Au and Ag assigned to A and B sites, as
follows. (i) A$_1$ZZ$3+5$ [Fig.~\ref{fig1}(e)]: Au in A and A$^\prime$
and Ag in B and B$^\prime$; (ii) A$_2$ZZ$3+5$ [Fig.~\ref{fig1}(f)]: Au
in A and B and Ag in A$^\prime$ and B$^\prime$; (iii) A$_3$ZZ$3+5$: Au
in A and B$^\prime$ and Ag in A$^\prime$ and B; and (iv) A$_4$ZZ$3+5$:
Ag in A and A$^\prime$ and Au in the B and B$^\prime$. In addition, we
considered two ZZ4 alloy structures, (i) A$_1$ZZ4
[Fig.~\ref{fig1}(c)]: with Au and Ag atoms alternating on both the A
(bottom) and B (top) sites; and (ii) A$_2$ZZ4: with Au atoms on the A
sites and Ag on the B sites. For all pure systems, we performed
calculations at fixed strain (length per atom), with full relaxation
of internal degrees of freedom. For the minima of each $E\times\ell$
curve, full relaxation was performed, allowing cell vectors to relax
until residual stresses were negligible. For the alloys, we only
performed full relaxations to obtain the unstrained minima of the
configurations described above.
\begin{table}
%\centering
\caption{Energy, length per atom, and electronic band gap (at the
level of GGA-DFT) of Au, Ag, and Ag$_{0.5}$Au$_{0.5}$ nanowires, at
the equilibrium geometry.}
\begin{tabular}{lccc}
\hline
\hline
Structure   &   Energy (eV/atom)   &  length (\AA/atom) &   Gap (eV)   \\
\hline
Ag - SQ3      & $- 1.85$       & 1.34   & 0.00      \\
Ag - ZZ4      & $- 1.99$       & 1.35   & 0.00      \\
Ag - ZZ3$+$5  & $- 2.08$       & 1.14   & 0.78      \\
\hline
Au - SQ3      & $- 2.35$      & 1.33  & 0.00       \\
Au - ZZ4      & $- 2.49$      & 1.34  & 0.00       \\
Au - ZZ3$+$5  & $- 2.61$      & 1.12  & 1.29       \\
\hline
A$_1$ZZ3$+$5  & $- 1.88$     & 1.14   & 1.85    \\
A$_2$ZZ3$+$5  & $- 1.77$     & 1.12   & 0.93    \\
A$_3$ZZ3$+$5  & $- 1.74$     & 1.13   & 1.06    \\
A$_4$ZZ3$+$5  & $- 1.69$     & 1.13   & 0.11    \\
A$_1$ZZ4      & $- 1.71$     & 1.33   & 0.59    \\
A$_2$ZZ4      & $- 1.65$     & 1.34   & 0.00    \\
\hline \hline
\end{tabular}
\end{table}

Figure~\ref{fig2} shows the $E\times\ell$ curves for each geometry,
for Au and Ag. All energies are referred to the energy of the isolated
atomic components. For Au (Ag), the minimum for the ZZ3$+$5 geometry
is 0.12 (0.09) eV/atom lower in energy than the one for the ZZ4
structure. Among the structures in Fig.~\ref{fig2}, the SQ3 geometry
shows the highest equilibrium energy for both metals. Our results for
the total energies, the equilibrium length, and the electronic band
gap, for each structure, are included in Table I. For the ZZ4, our
values are in good agreement with previously published
results.~\cite{liqin,ribeiro,springborg1,springborg2,portal}

Regarding the stability of the ZZ3$+$5, we believe our results to be
relevant in the context of the thinning process of the NWs in MBJE, as
discussed in Ref.~\onlinecite{cheng}, whenever the time scale of this
process allows for the stabilization of the successive local minima,
as the NW atomic density (and radius) decreases. In the atomic density
range we investigate in this work, the ZZ3$+$5 geometry is the most
stable, and should be a ``magic structure'' (in the language of
Ref.~\onlinecite{cheng}) adopted by the NW in the thinning process. We
can also look at the NW stability issue from the other end, i.e., at
the ultrathin limit, by identifying possible transitions between
locally stable NW structures, as the applied tension, hence the atomic
density, fluctuate. An upper bound for the barrier involved in the
ZZ4-to-ZZ3$+$5 transition is given by the energy difference between
the ZZ4 minimum and the crossing of the ZZ4 and ZZ3$+$5 curves in
Fig.~\ref{fig2}, plus the barrier for the transformation, when both
structures have the length per atom corresponding to the crossing
point. In Fig.~\ref{fig2}, we also show the $E\times\ell$ curve for a
non-planar variation of the ZZ4 geometry (NPZZ4), where small
non-planar distortions are allowed.  ZZ4 and NPZZ4 are very nearly
degenerate at the corresponding minima. Our upper-bound barrier for
the ZZ4-to-ZZ3$+$5 transformation is of the order of the
room-temperature thermal energy for Ag ($\sim$50~meV) and Au
($\sim$70~meV). In the case of Au, a lower barrier is obtained
starting from the NPZZ4, while in Ag a lower-barrier path starts from
the ZZ4 itself.
\begin{figure}
%\begin{figure}[htbp]
\includegraphics[width=8.5cm]{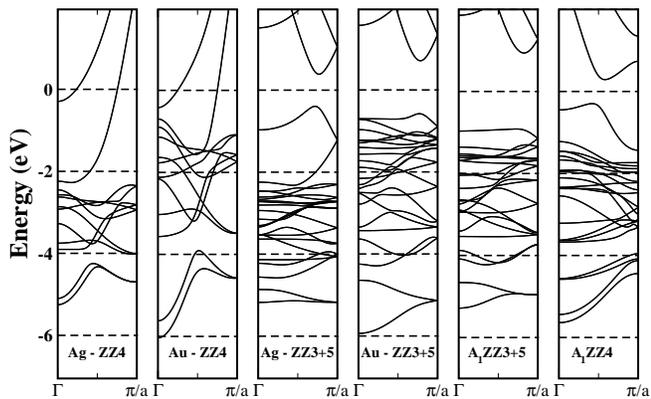}\\
\caption{Energy bands for Au, Ag, and Ag$_{0.5}$Au$_{0.5}$ nanowires
in ZZ4 and ZZ3$+$5 geometries.}
\label{fig3}
\end{figure}

The band structures of the ZZ4 and ZZ$3+5$ structures for Au and Ag
are shown in Fig.~\ref{fig3}(a)-(d). In both cases, the Fermi level
($E_f$) for the ZZ4 geometry does not occur at the $\pi/2a$ vector,
and the stability and semiconducting nature of the period-doubling
ZZ$3+5$ does not result from either a GPT or a Peierls instability. It
derives, rather, from intricate band-structure effects, which are
different between the two metals. We discuss first the band structure
of the ZZ4 geometries. We have analyzed in detail the
orbital-projected density of states resulting from the bands shown in
Fig.~\ref{fig3}. For both Ag and Au, the two bands crossing $E_f$ are
one almost-purely $s$ band, crossing near one quarter of the
Brillouin zone, and one hybridized $sd$ band crossing near the point
at three quarters of the zone, with predominantly $d$ character. Due
to the relativistic effect, the two metals differ in the nature of the
hybridized band: $s$ and $d$ levels are more intermixed in the case of
Au, while the $s$ levels are higher in energy with respect to $d$
levels, in the case of Ag. This explains the presence of a large
pseudo-gap region below the Fermi level in the case of Ag which
becomes very small in the case of Au, since the density of states
(DOS) of the $d$-dominant band peaks at much lower energies with
respect to the $s$ band in the case of Ag.

In the ZZ$3+5$ structure, the presence of two sites with different
coordinations causes the DOS peaks dominated by the threefold
(fivefold) sites to shift upwards (downwards), relative to the ZZ4
peaks. In a tight-binding language, the onsite terms are shifted due
the change in coordination. As a result, in the case of Ag, the gap is
between two nearly purely $s$ bands (plus the $p$ character from the
polarization orbitals). The top of the valence band (HOMO) is a
bonding $s$ band from the threefold atoms while the bottom of the
conduction band (LUMO) is an antibonding band from the fivefold
atoms. The bonding band from the fivefold sites is deeper in energy
than the HOMO, and the antibonding one from the threefold sites is
higher than the LUMO. In the case of Au, the HOMO band is a strongly
mixed $sd$ band with $d$ character from both types of sites (the
fivefold site is dominant), and $s$ character from the threefold
sites, while the LUMO is the same as in Ag, the antibonding $s$ band
from the fivefold sites.

For the alloys, the most stable A$_1$ZZ$3+5$ has Au in the threefold
sites and Ag in the fivefold sites, while in the high-energy
A$_4$ZZ$3+5$ Au and Ag switch places. This is due to stronger
relativistic effects in Au, that lead to its stronger tendency to form
low-coordinated structures. The gaps for these two alloys can be
obtained accurately from the HOMO and LUMO of the pure Au and Ag
structures. Since HOMO bands are dominated by threefold sites and LUMO
ones by fivefold sites, and HOMO and LUMO energies are deeper in Au
than in Ag, a larger gap is expected for A$_1$ZZ$3+5$, with Au at
threefold sites leading to a deeper HOMO, and Ag at fivefold sites
leading to a higher LUMO. The opposite effect would occur in
A$_4$ZZ$3+5$. This is indeed what we obtain, as shown in Table
I. Further, Table I shows a clear correlation between energy and band
gap for the alloy structures. Note that the insulating A$_1$ZZ4 alloy
has a lower energy than the A$_4$ZZ$3+5$.

In summary, the authors introduce a geometry for ultrathin Au, Ag, and
Au-Ag-alloy wires that is more stable than previously considered
planar geometries for these systems, by about 0.1 eV per atom. The
insulating nature of this structure, with gaps of 1.3~eV for Au and
0.8~eV for Ag, is not related to Peierls instabilities, resulting,
rather, from an interplay between geometric and electronic structure
effects.

%\section{ACKNOWLEDGMENTS}
The authors acknowledge support from the Brazilian agencies: FAPEMIG,
CAPES, CNPq, and Instituto do Mil\^enio de Nanotecnologia/MCT.\\


\begin{thebibliography}{99}

\bibitem{ohnishi} H.~Ohnishi, Y.~Kondo, and K.~Takayanagi, Nature
(London) {\bf 395}, 780 (1998).

\bibitem{yanson} A.~I.~Yanson, G.~Rubio-Bollinger, H.~E.~van~den~Brom,
N.~Agr\"{\i}t, and J.~M.~van~Ruitenbeek, Nature (London) {\bf 395},
783 (1998).

\bibitem{ugarte1} J.~Bettini, F.~Sato, P.~Z. Coura, S.~O.~Dantas,
D.~S.~Galv\~ao, and D.~Ugarte, Nature Nanotechnology {\bf 1}, 182
(2006).

\bibitem{ugarte2} L.~G.~C.~Rego, A.~R.~Rocha, V.~Rodrigues, and
D.~Ugarte, Phys. Rev. B {\bf 67}, Art.~No.~045412 (2003).

\bibitem{ugarte3} V.~Rodrigues, J.~Bettini, A.~R.~Rocha,
L.~G.~C.~Rego, and D.~Ugarte, Phys. Rev. B {\bf 65}, Art.~No.~153402
(2002).

\bibitem{liqin} L~Q.~Ke, M.~van~Schilfgaarde, T.~Kotani, and
P.~A.~Bennet, Nanotechnology {\bf 18}, Art.~No.~095709 (2007).

\bibitem{enomoto} A.~Enomoto, S.~Kurokawa, and A.~Sakai, Phys. Rev. B
{\bf 65}, Art.~No.~125410 (2002).

\bibitem{springborg1} M.~Springborg and P.~Sarkar, Phys. Rev. B
{\bf 68}, 045430 (2003)

\bibitem{springborg2} A.~M.~Asaduzzaman and M.~Springborg, Phys. Rev. B
{\bf 72}, 165422 (2005).

\bibitem{ribeiro} F.~J.~Ribeiro and M.~L.~Cohen, Phys. Rev. B {\bf
68}, 035423 (2003)

\bibitem{bahn} S.~R.~Bahn and K.~W.~Jacobsen, Phys. Rev. Lett. {\bf 87},
266101 (2001).

\bibitem{portal} D.~S\'{a}nchez-Portal, E.~Artacho, J.~Junquera,
P.~Ordej\'on, A.~Garc\'{\i}a, and J.~M.~Soler, Rev. Lett. {\bf 83},
3884 (1999).

\bibitem{smit} R.~H.~M.~Smit, C.~Untiedt, A.~I.~Yanson, and
J.~M.~van~Ruitenbeek, Phys. Rev. Lett. {\bf 87}, 266102 (2001).

\bibitem{gpt} I.~P.~Batra, Phys. Rev. B {\bf 42}, 9162 (1990).

\bibitem{cheng} D.~Cheng, W.~Y.~Kim, S.~K.~Min, T.~Nautiyal, and
K.~S.~Kim, Phys. Rev, Lett. {\bf 96}, Art.~No.~096104, 2006.

\bibitem{siesta} J.~M.~Soler, E.~Artacho, J.~D.~Gale, A.~Garcia,
J.~Junquera, P.~Ordejon, and D.~Sanchez-Portal,  J.~Phys-Condens.
Mat. {\bf 14}, 2745 (2002).

\bibitem{ks} W. Kohn, L. J. Sham. {\it Phys. Rev.} {\bf 140}, A1133
(1965).

\bibitem{pbe} J.~P.~Perdew, K.~Burke, and M.~Ernzerhof,
Phys. Rev. Lett. {\bf 77}, 3865 (1996).

\bibitem{tm} N.~Troullier and J.~L.~Martins, Phys. Rev. B {\bf 43},
1993 (1991).

\end{thebibliography}
\end{document}